# Non-Hermitian engineering of terahertz light using exceptional points in electrically tuneable collective light-matter interactions

M. Said Ergoktas[1,2], Sina Soleymani[3], Nurbek Kakenov[4], Thomas B. Smith[5], Gokhan Bakan[1,2], Kaiyuan Wang[1,2], Sinan Balci[6], Alessandro Principi[5], Kostya S. Novoselov[5], Sahin K. Ozdemir[3,7†], Coskun Kocabas[1,2,8†]

1. Department of Materials, University of Manchester, Manchester, M13 9PL, UK
2. National Graphene Institute, University of Manchester, Manchester, M13 9PL, UK
3. Department of Engineering Science and Mechanics, Pennsylvania State University, University Park, 16802 PA, USA
4. Department of Physics, Bilkent University, Ankara, Turkey
5. Department of Physics and Astronomy, University of Manchester, Manchester, M13 9PL, UK
6. Department of Photonics, Izmir Institute of Technology, Izmir, Turkey
7. Materials Research Institute, Pennsylvania State University, University Park, 16802 PA, USA
8. Henry Royce Institute for Advanced Materials, University of Manchester, Manchester M13 9PL, UK

† E-mail: sko9@psu.edu, coskun.kocabas@manchester.ac.uk

**The topological structure associated with the branchpoint singularity around an exceptional point (EP) provides new tools for controlling the propagation of electromagnetic waves and their interaction with matter [1–4]. To date, observation of EPs in light-matter interactions has remained elusive and has hampered further progress in applications of EP physics. Here, we demonstrate the emergence of EPs in the electrically controlled interaction of light with a collection of organic molecules in the terahertz regime at room temperature. We show, using time-domain terahertz spectroscopy, that the intensity and phase of terahertz pulses can be controlled by a gate voltage which drives the device across the EP. This fully electrically-tuneable system allows**

**reconstructing the Riemann surface associated with the complex energy landscape and provides a topological control of light by tuning the loss-imbalance and frequency detuning of interacting modes. We anticipate that our work could pave the way for new means of dynamic control on the intensity and phase of terahertz field, developing topological optoelectronics, and studying the manifestations of EP physics in the quantum correlations of the light emitted by a collection of emitters coupled to resonators.**

Light-matter interactions are fundamental to our understanding and observation of the universe as well as to a wide range of applications in the classical and quantum domains, including but not limited to sensing, imaging, light generation, information processing, and computation. The light component in these interactions is usually in the form of electromagnetic modes confined in a resonator while the matter component involves a single or a mesoscopic number of oscillators. Changing the number of oscillators coupled to a resonator is a legitimate route for achieving strong or weak light-matter coupling[5,6]; however, it is not a desirable one in many practical settings as it does not lend itself to a tuneable and finely controllable platforms that can enable studying both weak and strong coupling regimes, as well as transitions between them. The alternative is to keep the number of oscillators fixed while tuning the imbalance between the losses and coupling strength between the oscillators and the resonator such that the system is steered between the weak and strong coupling regimes[7]. Such a non-Hermitian engineering of the system inevitably gives rise to non-Hermitian degeneracies known as EPs, which coincide with the cross-over point between the weak and strong coupling regimes. EPs are strikingly different from the degeneracies of Hermitian systems, known as diabolic points (DPs)[8]: At a DP, only the eigenvalues coalesce but the corresponding eigenstates remain orthogonal. On the contrary, at an EP both the eigenvalues and the associated eigenvectors coalesce, modifying the energy landscape of the system drastically, thus resulting in reduced dimensionality and skewed topology. This, in turn, for example enhances the system's response to perturbations[9–15], modifies the local density of states leading to the enhancement of spontaneous emission rates[16–19], and leads to a plethora of counterintuitive phenomena such as loss-induced lasing[20,21], topological energy transfer[22], enhanced chiral absorption[23,24], linewidth enhancement in lasers[25,26], unidirectional emission in ring laser[27], and asymmetric mode switching[28], just to name a few.

In this Letter, we demonstrate, for the first time, the emergence of EPs in the interaction of light with a collection of organic molecules in the terahertz regime by non-Hermitian engineering. Differently from the prior demonstrations in optical[29,30], optomechanical[22,25,31], electronic[32], acoustic[33], and thermal systems[34], where EPs are observed in the interaction between two or more analogous modes, here, EPs emerge in the coupling of two modes with different physical origins. Using a fully electrically tuneable system, which helps finely tune the losses and vary the detuning between the frequencies of a THz resonator and the collective intermolecular vibrations of organic molecules, we have achieved (i) room temperature strong coupling in the terahertz, (ii) observed the emergence of EPs in the transition between the strong and weak coupling regimes, (iii) reconstructed the Riemann surface associated with the complex energy landscape of the system, and (iv) demonstrated the drastic effect of EPs on the intensity and phase of terahertz pulses. Our study provides the first direct evidence of EPs in the interaction of two disparate modes- one photonic mode and one phononic mode as the material excitation-and paves the way towards achieving and steering THz fields and their interactions with organic material by harnessing EPs. We note that this intriguing feature, that is the emergence of an EP in the coupling of light and matter modes (i.e., two modes with strikingly different physical origins) has been recently discussed in a theoretical work by J. Khurgin within the context of a polaritonic system where the possibility of adiabatic transform of the photons to excitons and vice versa was discussed[35].

The physical platform that we use in this study is a graphene-based tuneable THz resonator[36], which resembles a Gires–Tournois interferometer with the gate electrode forming the bottom reflective mirror and the graphene layer placed a distance away from it forming the tuneable top mirror (**Fig. 1a**). A non-volatile ionic liquid electrolyte layer is placed between the mirrors to achieve reversible gating of graphene by an applied voltage $V_1$ (i.e., effective gate voltage from the Dirac point), enabling an electrically tuneable reflectivity and hence resonator loss. The gate electrode (a 100 nm gold film evaporated on a 50 µm-thick Kapton film) is placed on a piezo stage driven by an applied voltage $V_2$, forming a moveable mirror that can be used to vary the cavity length and hence tune the resonance frequency (See Materials and Methods of Supporting Information for the details of the device fabrication). $\alpha$-lactose crystals that support collective intermolecular vibrations at $\omega_{vib}$ =0.53 THz[37] with a very narrow linewidth of $\gamma_{vib} = 0.023$ THz are embedded in the resonator to allow for studying the emergence of EPs in light-matter interactions (i.e., coupling between the resonator field and the α-lactose crystals) in the THz regime.

The dynamics of this coupled system, in which an ensemble of $N$ identical molecular vibrations of frequency $\omega_{vib}$ are coupled to a resonator mode of frequency $\omega_c$ with the same coupling strength $g$ are given by the complex eigenfrequencies $\omega_\pm = (\Delta + 2\omega_{vib})/2 - i(\Gamma + 2\gamma_{vib})/4 \pm \Omega/4$. The non-orthogonal eigenmodes are $|\psi_\pm\rangle \propto \begin{pmatrix} \omega_\pm \\ \sqrt{N}g \end{pmatrix}$. Here, $\Delta = \omega_c - \omega_{vib}$ is the frequency detuning and $\Gamma = \gamma_c - \gamma_{vib}$ represents the loss-imbalance between the molecular oscillators and the resonator, while $\gamma_c$ and $\gamma_{vib}$ are the decay rates of the resonator and molecular vibrations, respectively. Finally, $\Omega = \sqrt{16Ng^2 + (2\Delta + i\Gamma)^2}$ denotes the effective coupling strength between two systems. Analysis of this expression reveals that for $\Delta = 0$ (i.e., when the field is resonant with molecular vibrations) and $\sqrt{N}g > \Gamma/4$ (i.e., strong coupling regime), the complex eigenfrequencies exhibit splitting in their real parts while their imaginary parts remain coalesced. On the other hand, for $\sqrt{N}g < \Gamma/4$ (i.e., weak coupling regime) they exhibit splitting in their imaginary parts while the real parts coalesce, implying the modification of the decay rates of the eigenstates. For $\sqrt{N}g = \pm\Gamma/4$, the complex eigenfrequencies coalesce both in their real and imaginary parts, i.e. $\omega_\pm = \omega_{EP} = (\omega_c + \omega_{vib})/2 - i(\gamma_c + \gamma_{vib})/4$, and in their associated eigenmodes, i.e. $|\psi_\pm\rangle = |\psi_{EP}\rangle \propto \begin{pmatrix}\omega_{EP}\\\Gamma_{EP}\end{pmatrix}$ with $\Gamma_{EP} = \pm 4\sqrt{N}g$ implying the emergence of two EPs.

In our system (**Fig. 1a**), the knobs $V_1$ and $V_2$ are used to finely tune $\Gamma$ and $\Delta$, respectively, and allow us to observe the transition between the strong and weak coupling regimes through the EP. Plotting the complex energy landscape (i.e., real and imaginary parts of the complex eigenfrequencies $\omega_\pm$) as $V_1$ and $V_2$ are varied yields two intersecting Riemann sheets wrapped around a second-order EP right in the centre where the two complex eigenfrequencies of the system coalesce (**Fig. 1b**). Representing the eigenstates of the system on the Bloch sphere (**Fig. 1c**) allows us to monitor the evolution of the state of the system during the transition from weak to strong coupling through the EP. In the largely detuned or large loss-imbalance cases (i.e., $\Delta \to \infty$ or $\Gamma \gg \sqrt{N}g$, that is the limit of the uncoupled modes), the two supermodes of the system approach to the individual uncoupled electromagnetic mode (cavity photonic mode) and the matter mode (vibrational mode), which are located at the north and the south poles of the Bloch sphere, respectively. For $\Delta = 0$, varying $V_1$ and hence $\Gamma$ gradually shifts the supermodes from the poles distributing the supermodes across the cavity and the matter ($\alpha$-lactose crystals). The supermode close to the north pole mostly resides in the cavity (cavity-like mode) whereas the supermode close to south pole mostly resides in the matter (matter-like mode). With further tuning of $\Gamma$, the cavity-like mode $|c\rangle$ moves downward from the north pole, while the matter-like mode $|v\rangle$ moves upward from the south pole towards the equator. These modes then coalesce to the single mode $|\psi_{EP}\rangle$ on the equator at the critical value $\Gamma_{EP} = \pm 4\sqrt{N}g$ where dual EPs emerge.

We first experimentally test and confirm the effects of tuning knobs $V_1$ and $V_2$ (**Fig. 1a**) on the reflectivity of the empty THz resonator. It is clear that as the voltage $V_1$, which controls the cavity loss (and hence the loss imbalance Γ of the couple) is increased, the resonance frequency $\omega_c$ of the resonator remains intact, but the linewidth (proportional to the decay rate $\gamma_c$) of the cavity resonance becomes narrower and the resonance depth increases, approaching critical coupling (**Fig. 1d**). The second knob $V_2$ (cavity voltage), controls the length of the resonator and its resonance frequency $\omega_c$ by moving a piezo stage (hence the gate electrode) with respect to the graphene transistor with a resolution of less than 6 nm. This helps finely adjust the frequency detuning Δ. It is clearly seen that as $V_2$ is varied and hence, the resonance frequency $\omega_c$ of the THz resonator shifts with no significant variation in the resonance linewidth (**Fig. 1e**). Since these processes do not have any effect on the vibrational frequency and decay rate of the molecules (**Fig. S1**), knobs $V_1$ and $V_2$ effectively control the 2-dimensional parameter space of Δ and Γ. We observed a tunability of around $\pm 25\ GHz$ in Δ and $100 GHz$ in Γ when $V_1$ and $V_2$ were increased from 0 to 1 V (**Fig. 1f**). As a result, the knobs enable non-Hermitian engineering of the light-matter interaction between the THz resonator field and the collective intermolecular vibrations and allow us to map the complex energy landscape of the hybrid system.

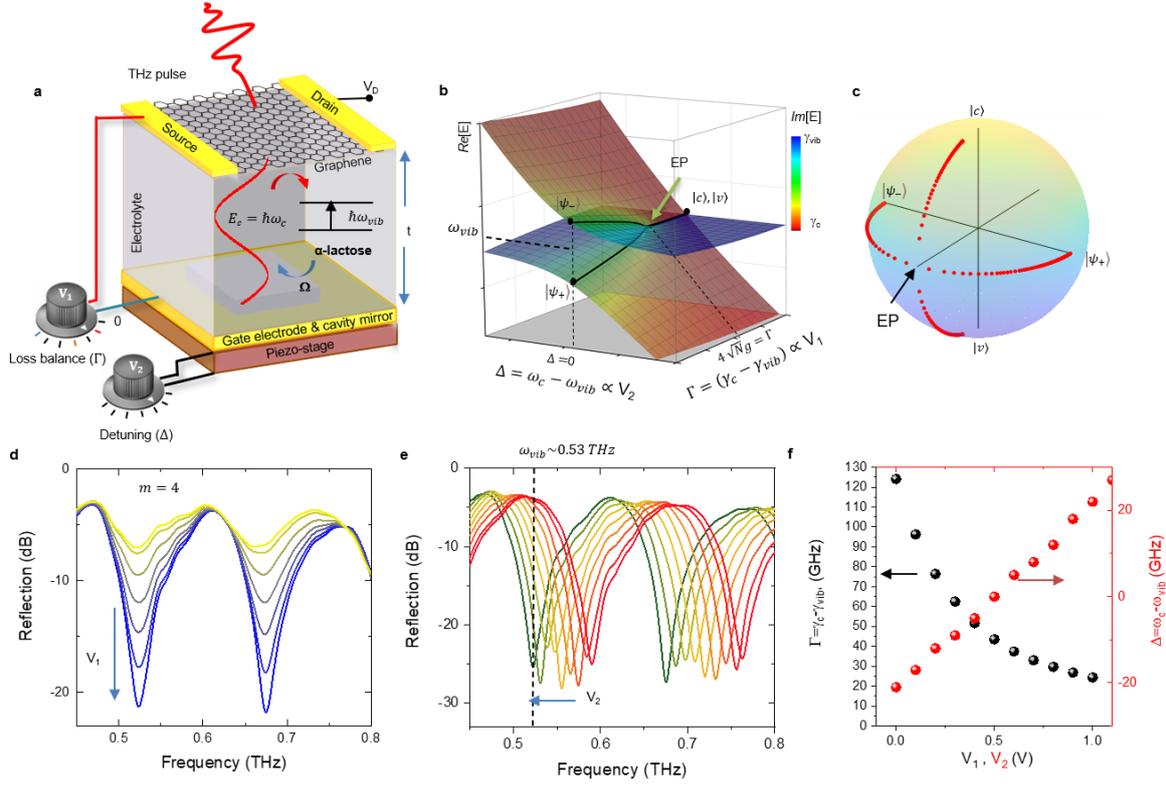

**Figure 1. Electrically tuneable EP device. a**, Schematic representation of the electrolyte-gated graphene transistor embedded with lactose microcrystals. The tuneable coupling between the resonator mode $E_c = \omega_c + i\gamma_c$ and the intermolecular vibrations of lactose crystals $E_{vib} = \omega_{vib} + i\gamma_{vib}$ forms an electrically tuneable two-parameter framework to realize EP devices. The gate voltage $V_1$ controls the loss imbalance, $\Gamma$ between the cavity and intermolecular vibrations by tuning the charge density on graphene and $V_2$ controls the detuning frequency, $\Delta$ by changing the cavity size. **b,** Riemann surface obtained using numerical simulations shows the complex energy eigenvalues of the device plotted on the two-parameter voltage space defined by $V_1$ and $V_2$. EP emerges when the coupling strengths compensates the loss imbalance, $\sqrt{N}g = \pm\Gamma/4$, when the cavity field and the intermolecular vibrations are on resonant, $\Delta = \omega_c - \omega_{vib} = 0$. **c,** Visualization of the evolution of the supermodes of the coupled system on a Bloch sphere as the gate voltage $V_1$ is varied (loss imbalance $\Gamma$ is tuned). The azimuthal angle on the sphere indicates the relative phase and the polar angle represents the relative intensity of the uncoupled cavity (photon mode) and the collective molecular vibrations (matter mode) represented by the eigenmodes $|c\rangle$ and $|v\rangle$, respectively. **d, e,** THz reflection spectrum of an empty cavity showing the dependence of the cavity mode $|c\rangle$ on $V_1$ and $V_2$, respectively. **f**, Voltage dependence of the loss imbalance $\Gamma$ and detuning $\Delta$ of the system.

Next, we perform time-domain THz spectroscopy to demonstrate tuneable transition between the weak and strong coupling regimes through an EP. We first tuned $V_2$ to have $\Delta =$

0, and then varied the gate voltage $V_1$, which controls the loss imbalance of the couples. As $V_1$ is increased, the formation of the characteristic polariton branching around $\omega_{vib}$ is clearly observed in the reflectivity map of the device (**Fig. 2a**). This branching takes place at two symmetric EPs $V_{EP} = \pm 0.2V$ due to ambipolar electrical conduction of graphene. A cross-section of this reflectivity map around one of these EPs reveals the transition from a split mode spectrum (i.e., strong coupling regime) to a coalesced mode spectrum (i.e., weak coupling regime) through the EP (**Fig. 2b**). The transition between these two regimes as $V_1$ is varied can be attributed to the variation of the optical conductivity of graphene and the corresponding cavity decay time (**Fig. 2c**). This dependence on $V_1$ clarifies our ability to control loss-imbalance between the couples through the control of the resonator losses.

Experiments with different cavity modes (from $m = 2$ to $m = 9$, adjusted by tuning the cavity size) satisfying $\Delta = 0$ reveal that the transition from the split modes to coalesced modes occurs at different $V_1$ voltages for different cavity modes (**Fig. 2d**): The higher is the mode number $m$, the smaller is the required gate voltage $V_1$ to arrive at the EP. This behaviour may be attributed to (i) the smaller number $N$ of molecules interacting with the cavity fields and thus the reduced effective coupling strength at higher $m$ (i.e., modes with higher $m$ have smaller spatial overlap with the molecule ensemble, thus a smaller number of molecules contributes to the process) or (ii) the smaller $\gamma_c$ of higher order modes and thus smaller initial loss imbalance between the couples. As a result, the amount of additional loss-imbalance required to satisfy the EP condition $\sqrt{N}g = \Gamma/4$ is smaller for higher order cavity modes, implying that modes with higher $m$ require smaller gate voltage $V_1$ to reach EP. Since the EP is a singularity point in the 2-parameter space, we have finely tuned $\Gamma$ and $\Delta$ via the knobs $V_1$ and $V_2$ for a fixed mode $m$ and reconstructed the Riemann surface associated with the complex energy landscape of the system (**Fig. 2e**). The topology of two intersecting Riemann sheets centred around an EP is clearly seen (**Figs. 1b, 2e**).

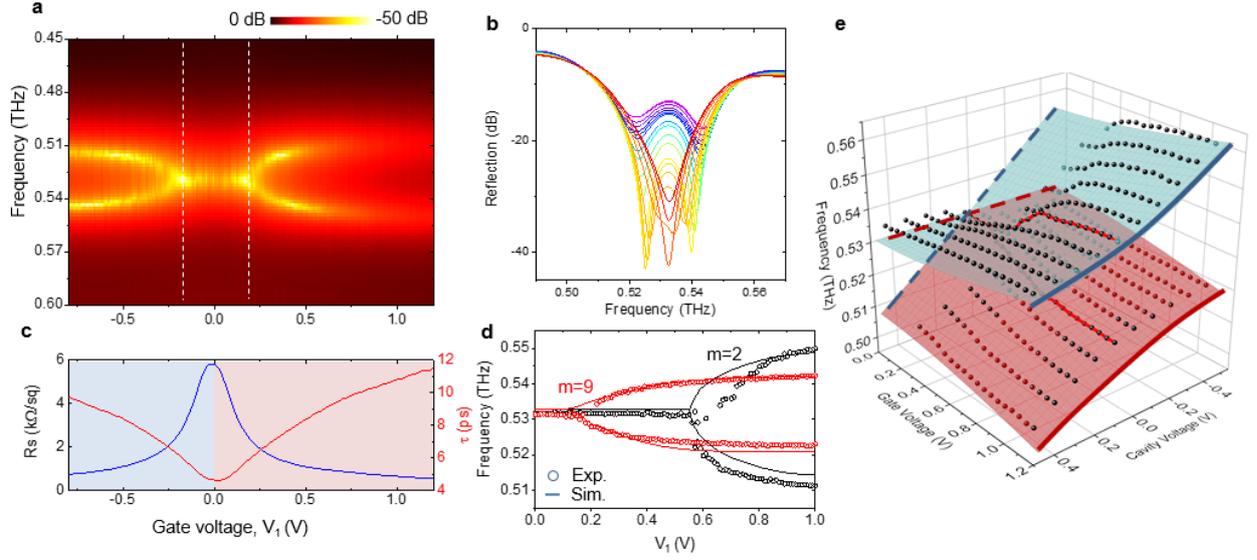

**Figure 2. Spectroscopic characterization of the EP device**. **a**, **b,** Reflectivity map and the spectra of the device showing the transition from the weak (coalesced modes) to the strong coupling (split modes) regimes through an EP as $V_1$ is varied (Γ is tuned) at $V_2$ satisfying $\Delta = 0$. Owing to the ambipolar conduction of graphene, the device goes through two EPs at $V_{EP1} = -0.2V$ (electron doping) and $V_{EP2} = 0.2V$ (hole doping). **c**, Sheet resistance of graphene and the cavity decay time plotted against the gate voltage. Increasing the gate voltage enhances the THz reflectivity of the graphene mirror leading to a longer cavity decay time. **d**, Position of the EP and the amount of splitting vary with the mode number $m$. EPs emerge at smaller gate voltages for higher $m$. **e**, Experimentally obtained (black dotted) and calculated (blue and red sheets) Riemann surfaces showing the real part of complex eigenvalues of the device in the voltage-controlled parameter space.

Next, we investigate the dynamical control of EP and its effect on the intensity and the phase of the reflected THz light. For this purpose, we prepare the system at $\Delta = 0$ and dynamically modulate the loss-imbalance Γ by applying a periodic square-wave gate voltage $V_1$. The time dependent reflection spectra clearly show periodic splitting and coalescence of the modes (**Fig. 3a**). The system gradually transits from the coalesced modes around $0.535\ THz$ to split modes with a splitting of around $40\ GHz$ in ~$0.2\ s$ after the gate voltage is set to ON state. We recorded the intensity (**Fig. 3b**) and the phase (**Fig. 3c**) of the reflected THz pulse from the device at different time delays after the ON signal is applied. We must point out that the measured phase depends on the reference plane, however, the phase difference is uniquely defined. We observe a phase accumulation of $0, 2\pi,$ and $4\pi$ across the free spectral range of the resonator during the transition through the EP. This geometrical (i.e., Berry) phase is the result of the topology of the Fresnel reflectivity $r(\omega)$. Here the topological invariant is the winding number $n = \frac{1}{2\pi i}\oint \frac{dr}{r}$ of the complex Fresnel reflectivity around the perfect absorption singularity ($r = 0$; critical coupling) where the reflection phase is undefined[38]. Calculated reflection (**Fig. 3d**) for our device at three different sheet resistances reveals three topologically different reflectivity identified by winding numbers $n = 0, 2$ and $1$ and the associated Berry phases of $0, 2\pi$ or $4\pi$, respectively, agreeing with the phases measured in the experiments (**Fig. 3c**). These results provide the first direct evidence for the electrically switchable reflection topology.

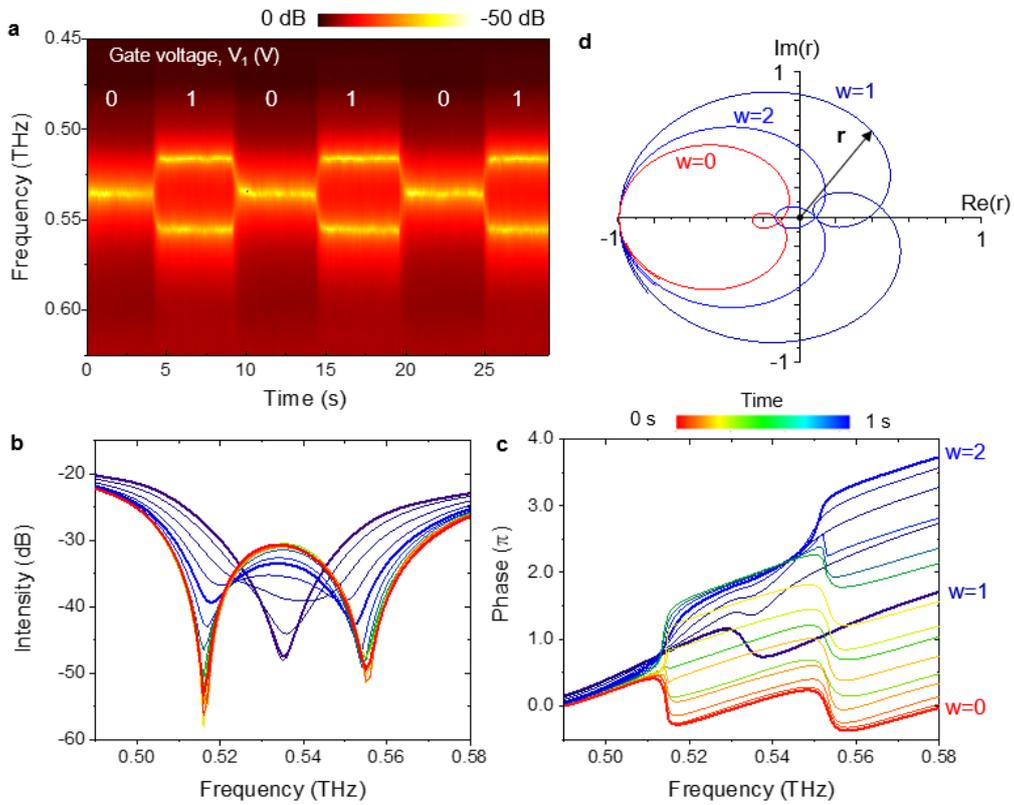

**Figure 3. Higher winding number topological switching around an EP. a**, Time dependent variation of the reflection spectrum of the device under a periodic square-wave gate voltage. **b, c,** Variation of the intensity and the phase of the reflected THz pulse from the device recorded at different time delays after the gate voltage is applied. **d**, Complex representation of the Fresnel reflection calculated for the device showing topologically different states at sheet resistances $R_s = 400, 700$ and $5000\,\Omega$ with winding numbers 0, 2 and 1. The effective gate voltage controls the transition between these states resulting in geometric phase accumulation of 0, $2\pi$ or $4\pi$ in good agreement with the measurement results in **c**.

One of the most intriguing features of an EP is the exchange of the eigenstate when it is adiabatically encircled. This is in contrast to encircling a DP in Hermitian systems where the eigenstate acquires a geometric phase, and no state flip takes place. While one loop around the EP flips the eigenstate, only the second loop returns the system to its initial state apart from a Berry phase $\pi$. State flip when encircling EPs have been experimentally demonstrated using static measurements of eigenvectors and eigenfrequencies in microwave cavities[39], optical resonators[40,] exciton-polariton systems[41], and acoustic systems[42]. Finally, we probe the dynamics of our system when it is steered on cyclic paths encircling an EP by tuning $\Gamma$ and $\Delta$ using the knobs $V_1$ and $V_2$. This is possible in our system because the two finely controlled knobs are independent. We note again that different from previous experimental studies probing static encircling of EPs, the EP in our study emerges in the interaction of two disparate modes. By varying $V_1$ and $V_2$ in steps of 25 mV such that an EP is encircled in the clockwise or counter-clockwise directions, we monitor how the final state of the system is affected by the encircling process. In order to do this we defined a loop by the points $\{\Delta_{max}, \Gamma_{min}\}$, $\{\Delta_{max}, \Gamma_{max}\}, \{\Delta_{min}, \Gamma_{max}\}, \{\Delta_{min}, \Gamma_{min}\}$ returning back to $\{\Delta_{max}, \Gamma_{min}\}$ after ~ 20 s. Similarly, in the parameter space of $V_1$ and $V_2$, the loop is defined by the corresponding voltage points as $\{V_{2max}, V_{1min}\}$, $\{V_{2max}, V_{1max}\}$, $\{V_{2min}, V_{1max}\}$, $\{V_{2min}, V_{1min}\}$ returning back to $\{V_{2max}, V_{1min}\}$. When we choose a control loop that does not enclose the EP, the system returns to the same state at the end of the loop (**Fig. 4a**). This is regardless of whether the loop is clockwise or counter-clockwise. In contrast, when the loop encircles the EP, we observe that a trajectory starting on one of the Riemann sheets ends on the other sheet (**Fig. 4b**) resulting in eigenstate exchange (state flip): $|\psi_+\rangle \to |\psi_-\rangle$ and $|\psi_-\rangle \to |\psi_+\rangle$. To get more insight on these dynamics, we illustrate the evolution of the eigenstates of the system on Bloch spheres for closed loops that do not encircle (**Fig. 4c**) and that encircle the EP (**Fig. 4d**). When the system is initially in the state $|\psi_+\rangle = (|c\rangle + |v\rangle)/\sqrt{2}$, which is the equal to the superposition of the cavity $|c\rangle$ and vibrational $|v\rangle$ modes, the final state after a closed loop encircling the EP becomes $|\psi_-\rangle = (|c\rangle - |v\rangle)/\sqrt{2}$ which is orthogonal to the initial state $|\psi_+\rangle$. A second loop around the EP brings the system back to its initial state $|\psi_+\rangle$ apart from a geometrical phase. As it is seen in the Bloch sphere (**Fig. 4d**), these two loops around the EP cut the Bloch sphere directly in half and correspond to a solid angle of $2\pi$, which in turn implies that the acquired geometrical phase is $\pi$ (i.e., the geometrical phase is the half of the solid angle enclosed by the curve connecting the initial and final states).

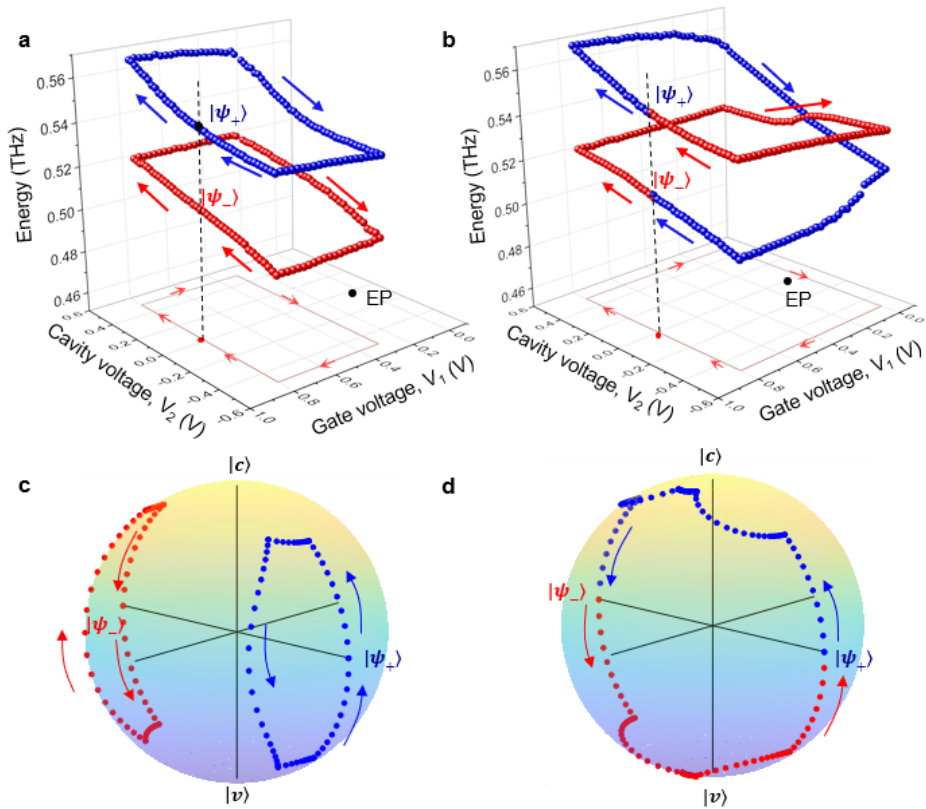

**Figure 4. Voltage-controlled encircling of EP. a, b,** Evolution of the energy of the coupled system along the trajectories traced by varying the voltages $V_1$ and $V_2$ in small steps. **a,** A trajectory starting on one of the Riemann sheets stays on the same sheet if it does not encircle the EP. **b,** A trajectory starting on one of the Riemann sheets ends on the other sheet (state

exchange) if it encircles the EP. **c, d,** Evolution of eigenstates of the system on the Bloch sphere for the trajectories shown in **a** and **b**, respectively.

In conclusion, we demonstrate a non-Hermitian optical device to study EP in the collective interaction of vibrational modes of organic molecules with THz field. Using fully electrically tuneable independent knobs, we have dynamically steered the system through an EP which enables electrical control on reflection topology. Our results will pave the way towards topological control of light-matter interactions around an EP, with potential applications ranging from quantum information science and topological optoelectronic devices to topological control of physical and chemical processes.

**Acknowledgement:** C.K. acknowledges support from European Research Council through ERC-Consolidator Grant (grant no 682723, SmartGraphene) and S.K.O. acknowledges support from AFOSR (Award no. FA9550-18-1-0235, FA9550-21-1-0202).

**Author contributions:** M.S.E., C.K., and S.K.O. conceived the idea. M.S.E. synthesized the graphene samples and fabricated the devices. M.S.E. and C.K. performed the experiments. S.S. and S.K.O. performed the simulation and developed the theory. N.K. helped with the measurements. M.S.E., S.K.O. and C.K. analysed the data and wrote the manuscript with input from all the authors. All authors discussed the results and contributed to the scientific interpretation as well as to the writing of the manuscript.

**Additional information:** Authors declare no competing financial interests.